\documentclass[twoside,10pt]{article}
\usepackage{amsthm,amsmath,amssymb,amscd,enumerate}
\usepackage{amsfonts}
\usepackage{graphicx}
\usepackage{fancyhdr}
\begin{document}


\begin{center} {\Large\bf The emergence of the Virasoro and $w_{\infty}$  algebras through the renormalized  powers of quantum white noise}
\vskip 1cm {\bf   Luigi Accardi$^{1}$, Andreas Boukas$^{2}$}\\
\
\\$^{1}$ Centro Vito Volterra\\
Universit\`{a} di Roma Tor Vergata \\
via Columbia  2, 00133 Roma, Italia\\
e-mail: accardi@volterra.mat.uniroma2.it\\
\
\\$^{2}$ Department of Mathematics and Natural Sciences\\
American College of Greece\\
Aghia Paraskevi, Athens 15342, Greece\\
e-mail: andreasboukas@acgmail.gr\\
\end{center}

\begin{abstract} We introduce a new renormalization for the powers of the Dirac  delta function.  
We show that this new renormalization leads to a second quantized version of the Virasoro 
sector $w_{\infty}$ of the extended conformal algebra with infinite symmetries $W_{\infty}$ 
of Conformal Field Theory ( \cite{4a}-\cite{4d}, \cite{ketov}, \cite{7}, \cite{8}). In particular
we construct a white noise (boson) representation of the $w_{\infty}$ generators and commutation relations and of their second quantization.  
\end{abstract}

\numberwithin{equation}{section}
\newtheorem{theorem}{Theorem}
\newtheorem{lemma}{Lemma}
\newtheorem{proposition}{Proposition}
\newtheorem{corollary}{Corollary}
\newtheorem{example}{Example}
\newtheorem{algorithm}{Algorithm}
\newtheorem*{main}{Main~Theorem}
\newtheorem{notation}{Notation}
\newtheorem{remark}{Remark}
\newtheorem{definition}{Definition}

\section{Introduction}

Classical (i.e It\^{o}  \cite{5}) and quantum (i.e Hudson-Parthasarathy \cite{6}) stochastic calculi were unified by Accardi, Lu and Volovich in \cite{4}  in the framework of Hida's white noise theory by expressing the fundamental noise processes in terms of the Hida white noise functionals $b_t$ and $b_t^{\dagger}$ defined as follows:  Let $L^2_{sym}({\mathbb{R}}^n)$ denote the space of square integrable functions on ${\mathbb{R}}^n$ which are symmetric under permutation of their arguments  and let ${\cal F}:=\bigoplus^\infty_{n=0}L^2_{sym}({\mathbb{R}}^n)$ where if $\psi:=\{\psi^{(n)}\}^\infty_{n=0}\in {\cal F}$, then  $\psi^{(0)}\in{\bf C}$,  $\psi^{(n)}\in L^2_{sym}({\mathbb{R}}^n)$ and 

\[
\|\psi\|^2=\|\psi(0)\|^2+\sum^\infty_{n=1}\int_{{\mathbb{R}}^n}|\psi^{(n)}(s_1,\dots,s_n)|^2ds_1\dots ds_n
\]

The subspace of vectors $\psi=\{\psi^{(n)}\}^\infty_{n=0}\in {\cal F} $ with $\psi^{(n)}=0$ for  all but finitely many $n$ will be denoted by ${\cal D}_0$. Denote by $S\subset L^2({\mathbb{R}}^n)$ the Schwartz space of smooth functions decreasing at infinity faster than any polynomial and let ${\cal D}$  be the set of all $\psi\in {\cal F}$ such that $\psi^{(n)}\in S$ and $\sum^\infty_{n=1}n\,|\psi^{(n)}|^2<\infty$. For each  $t\in{\mathbb{R}}$ define the linear operator $b_t: {\cal D}\to {\cal F}$ by

\[
(b_t\psi)^{(n)}(s_1,\dots,s_n):=\sqrt{n+1}\psi^{(n+1)}(t,s_1,\dots,s_n)
\]

and the operator valued distribution  $b^{\dagger}_t$ by

\[
(b^{\dagger}_t\psi)^{(n)}(s_1,\dots,s_n):=\frac{1}{\sqrt n}\,\sum^n_{i=1}\delta
(t-s_i)\psi^{(n-1)}(s_1,\dots,\hat s_i,\dots,s_n)
\]

where $\delta$ is the Dirac delta function and  $\,\hat{}\,$ denotes omission of the corresponding variable.  The  white noise functionals satisfy the Boson commutation relations

\[
[b_t,b_s^{\dagger}]=\delta(t-s)
\]

\[
\lbrack b_t^{\dagger},b_s^{\dagger}\rbrack=[b_t,b_s]=0
\]

and the duality relation

\[
(b_s)^*=b_s^{\dagger}
\]

Letting  $\mathcal{H}$ be a test function space we define for  $f\in\mathcal{H}$ and $n,k\in\{0,1,2,...\}$ the sesquilinear form on ${\cal D}_0$ 

\[
B_k^n(f):=\int_{\mathbb{R}}\,f(t)\,{b_t^{\dagger}}^n\,b_t^k\,dt
\]

i.e for $\phi$, $\psi$ in ${\cal D}_0$   and $n,k\geq0$ 

\[
<\psi, B^n_k(f)\,\phi >=\int_{\mathbb{R}}\,f(t)\,<b_t^n\,\psi, b_t^k\,\phi>\,dt
\]

with involution

\[
\left(B_k^n(f)\right)^{*} = B_n^k(\bar f)
\]

and with

\[
B^0_0 (\bar g f)=\int_{\mathbb{R}}\, \bar g(t)\,f(t)\,dt=<g,f>
\]
The Fock representation is characterized by the existence of a unit vector $\Phi$, 
called the Fock vacuum vector, cyclic for the operators $B_n^k(\bar f)$ and satisfying:
 \begin{equation}\label{dffckvac}
B^0_k\Phi= B^h_k\Phi =0 \qquad ; \qquad \forall k>0 \ ;  \  \forall h\geq 0
\end{equation}

It is not difficult to prove that, if the Fock representation exists, it is uniquely 
characterized by the two above mentioned properties.

In \cite{1} it was proved that for all $t,s\in\mathbb{R_+}$ and $n,k,N,K\geq0$, one has:

 \begin{equation}\label{1}
[{b_t^{\dagger}}^nb_t^k,{b_s^{\dagger}}^Nb_s^K]=
\end{equation}

\[
\epsilon_{k,0}\epsilon_{N,0}\sum_{L\geq 1} \binom{k}{L}N^{(L)}\,{b_t^{\dagger}}^{n}\,{b_s^{\dagger}}^{N-L}\,b_t^{k-L}\,b_s^K\,\delta^L(t-s)
\]

\[
-\epsilon_{K,0}\epsilon_{n,0}\sum_{L\geq 1} \binom{K}{L}n^{(L)}\,{b_s^{\dagger}}^{N}\,{b_t^{\dagger}}^{n-L}\,b_s^{K-L}\,b_t^k\,\delta^L(t-s)
\]

where

\[
 \epsilon_{n,k}:=1-\delta_{n,k}
\]

$\delta_{n,k}$ is Kronecker's delta and the decreasing factorial powers $x^{(y)}$ are defined by

\[
x^{(y)}:=x(x-1)\cdots (x-y+1)
\]

with $x^{(0)}=1$.   In order to consider higher powers of  $b_t$ and $b_t^{\dagger}$, the renormalization

\begin{equation}
\delta^l(t)=c^{\,l-1}\,\delta(t),\,\,\,\,\,l=2,3,....\label{f1}
\end{equation}

where $c>0$ is an arbitrary constant, was introduced in  \cite{4}. Then (\ref{1}) becomes

 \begin{equation}
[{b_t^{\dagger}}^nb_t^k,{b_s^{\dagger}}^Nb_s^K]=\label{1a}
\end{equation}

\[
\epsilon_{k,0}\epsilon_{N,0}\sum_{L\geq 1} \binom{k}{L}N^{(L)}\,c^{\,L-1}\,{b_t^{\dagger}}^{n}\,{b_s^{\dagger}}^{N-L}\,b_t^{k-L}\,b_s^
K\,\delta(t-s)
\]

\[
-\epsilon_{K,0}\epsilon_{n,0}\sum_{L\geq 1} \binom{K}{L}n^{(L)}\,c^{\,L-1}\,{b_s^{\dagger}}^{N}\,{b_t^{\dagger}}^{n-L}\,b_s^{K-L}\,b_t^k\,\delta(t-s)
\]

Multiplying both sides  of (\ref{1a}) by test functions $f(t)\bar g (s)$ and formally integrating the resulting identity (i.e. taking  $\int\int\, \dots \,dsdt$), we obtain the following 
commutation relations for the Renormalized Powers of Quantum White Noise (RPQWN)

\begin{equation}\label{2}
[B^N_K(\bar g),B^n_k(f) ]
\end{equation}

\[
=\sum_{L= 1}^{K \wedge n}\, b_L(K,n)\,B^{N+n-L}_{K+k-L}(\bar g f)-\sum_{L= 1}^{k \wedge N }\,b_L(k,N)\,B^{N+n-L}_{K+k-L}(\bar g f)
\]

\[
=\sum_{L= 1}^{ (K \wedge n) \vee ( k \wedge N)  }\, \theta_L (N,K;n,k)
\,c^{L - 1}\,B^{N + n - L}_{K + k - L} (\bar g f)
\]

where
\begin{equation}\label{dfbLnm}
b_x(y,z):=\epsilon_{y,0}\,\epsilon_{z,0}\, \binom{y}{x}\,z^{(x)}\,c^{x-1}
\end{equation}

and for $n,k,N,K\in \{0,1,2,...\}$

\begin{equation}
\theta_L(N,K;n,k):=\epsilon_{K, 0}\,\epsilon_{n, 0}\,\binom{K}{ L}\, n^{(L)} - \epsilon_{k, 0}\,\epsilon_{N, 0} \,\binom{k}{L} \, N^{(L)}\label{2a}
\end{equation}

with

\[
\sum_{L= 1}^{ (K \wedge n) \vee ( k \wedge N)  }=0
\]

if $ (K \wedge n) \vee ( k \wedge N) =0$.  In what follows we will use the notation

\begin{equation}
B^n_k:=B^n_k(\chi_I)\label{not}
\end{equation}

whenever $I\subset \mathbb{R}$ with $\mu (I) < +\infty$ is fixed.  Moreover, to simplify the notations, we will use the same symbol for the generators of the RPQWN  algebra and for their images in a given representation. As above, we denote by $\Phi$ the Fock vacuum vector with  $b_t\,\Phi=0$ and $\langle\Phi,\Phi\rangle=1$.  It  was proved in \cite{1} that, with commutation relations (\ref{2}), the $B^n_k$ do not admit a common Fock space representation.  
The main counter-example is that if a common Fock representation of the $B^n_k$ existed, 
one should be able to define inner products of the form

\[
<(a\,B^{2n}_0(\chi_I)+b\,(B_0^n(\chi_I))^2)\Phi, (a\,B^{2n}_0(\chi_I)+b\,(B_0^n(\chi_I))^2)\Phi>
\]

where  $a,b\in \mathbb{R}$ and  $I$  is an arbitrary interval of finite measure $\mu(I)$. Using the notation $<x>=<\Phi,x\,\Phi>$  this amounts to the positive semi-definiteness of the quadratic form

\[
a^2\,<B^0_{2n}(\chi_I)\,B^{2n}_0(\chi_I)>+2\,a\,b\,<B^0_{2n}(\chi_I)(B^n_0(\chi_I))^2>
\]

\[
+\,\,b^2\,< (B^0_n(\chi_I))^2\,(B^n_0(\chi_I))^2>
\]

or equivalently of the $(2\times 2)$ matrix

\[
A=\left[
\begin{array}{ccc}
 <B^0_{2n}(\chi_I)\,B^{2n}_0(\chi_I)> &&<B^0_{2n}(\chi_I)\,(B^n_0(\chi_I))^2>\\
<B^0_{2n}(\chi_I)\,(B^n_0(\chi_I))^2> &&< (B^0_n(\chi_I))^2\,(B^n_0(\chi_I))^2>
\end{array}
\right]
\]

Using the commutation relations (\ref{2}) we find that 

\[
A=\left[
\begin{array}{ccc}
  (2n)!c^{2n-1}\mu (I)&&
(2n)!c^{2n-2}\mu (I)
\\
\\
(2n)!c^{2n-2}\mu (I) &&2 (n!)^2c^{2n-2} \mu (I)^2 +\left((2n)!-2(n!)^2\right) c^{2n-3} \mu (I)
\end{array}
\right]
\]

The matrix $A$ is  symmetric, so it is positive semi-definite only if its minors are non-negative. The minor determinants of $A$ are

\[
d_1= (2n)!c^{2n-1}\mu (I)\geq0
\]
and
\[
d_2= 2c^{4(n-1)}\mu (I)^2 (n!)^2 (2n)!(c\,\mu (I)-1) \geq 0   \Leftrightarrow
\mu (I) \geq \frac{1}{c}.
\]

Thus the interval $I$ cannot be arbitrarily small. The counter-example was extended in \cite{2}  to the $q$-deformed case

\[
b_t\,b_s^{\dagger}-q\,b_s^{\dagger}\,b_t=\delta(t-s) 
\]

A stronger no-go theorem, which establishes the impossibility of a Fock representation of any 
Lie algebra containing $B^n_0$ for any $n\geq 3$ and satisfying commutation relations  
(\ref{2}), can be proved using the following results.

\begin{lemma}\label{lem11} Let $n\geq3$ and define
\[
C_1(n):=[B^0_n,B^n_0 ]
\]
and for $k\geq2$
\[
C_k(n):=[B^0_n,C_{k-1}(n) ].
\]
Then
\begin{equation}
C_3(n)=\beta (n)\,B^0_{2n} +N(n) \label{55}  
\end{equation}
where, in the notation (\ref{dfbLnm}), $\beta (n) \in \mathbb{R}-\{0\}$ is given by
\begin{equation}
\beta (n):=  \sum_{L_1=1}^{n-1}\,\sum_{L_2=1}^{n-L_1}\, 
b_{L_1}(n,n) \,b_{L_2}(n,n-L_1) \,b_{n-(L_1+L_2)}(n,n-(L_1+L_2))\label{66} 
\end{equation}
and $N(n)$ is a sum of operators  given by
 \begin{equation}
N(n):=\sum_{L_1=1}^{n-1}\,\sum_{L_2=1}^{n-L_1}\,\sum_{L_3= 1}^{n-(L_1+L_2)}\, b_{L_1}(n,n) \,b_{L_2}(n,n-L_1) \label{77} 
\end{equation}
\[
\times b_{L_3}(n,n-(L_1+L_2))B^{n-(L_1+L_2+L_3)}_{3n-(L_1+L_2+L_3)}
\]
with adjoint

\begin{equation}
N(n)^*:=\sum_{L_1=1}^{n-1}\,\sum_{L_2=1}^{n-L_1}\,\sum_{L_3= 1}^{n-(L_1+L_2)}\, b_{L_1}(n,n) \,b_{L_2}(n,n-L_1) \label{88}
\end{equation}

\[
\times \,b_{L_3}(n,n-(L_1+L_2))B_{n-(L_1+L_2+L_3)}^{3n-(L_1+L_2+L_3)} 
\]

where the triple summations in  (\ref{77}) and (\ref{88})  are over all $L_1,L_2,L_3$ such that $L_1+L_2+L_3\neq n $.

\end{lemma}

\begin{proof} The commutation relations (\ref{2}) imply that:

\[
C_1(n)=[B^0_n,B^n_0 ]=\sum_{L_1= 1}^n\, b_{L_1}(n,n)\,B^{n-L_1}_{n-L_1}
\]

and

\[
C_2(n)=[B^0_n, C_1(n)]=\sum_{L_1= 1}^n\, b_{L_1}(n,n)\,[B^0_n,B^{n-L_1}_{n-L_1}]
\]

\[
=\sum_{L_1= 1}^n\, \sum_{L_2= 1}^{n-L_1}\, b_{L_1}(n,n)\,b_{L_2}(n,n-L_1)\,B^{n-(L_1+L_2)}_{2n-(L_1+L_2)}
\]

\[
=\sum_{L_1= 1}^{n-1}\, \sum_{L_2= 1}^{n-L_1}\, b_{L_1}(n,n)\,b_{L_2}(n,n-L_1)\,B^{n-(L_1+L_2)}_{2n-(L_1+L_2)}
\]

since  $[ B^0_n , B^{n-L_1}_{n-L_1} ]=0$ for $L_1=n$, and finally

\[
C_3(n)=[B^0_n, C_2(n)]
\]

\[
=\sum_{L_1= 1}^{n-1}\, \sum_{L_2= 1}^{n-L_1}\, b_{L_1}(n,n)\,b_{L_2}(n,n-L_1)\,[B^0_n ,B^{n-(L_1+L_2)}_{2n-(L_1+L_2)}]
\]

\[
=\sum_{L_1= 1}^{n-1}\, \sum_{L_2= 1}^{n-L_1}\,\sum_{L_3= 1}^{n-(L_1+L_2)}\, b_{L_1}(n,n)\,b_{L_2}(n,n-L_1)
\]

\[
\times b_{L_3}(n,n-(L_1+L_2))\,B^{n-(L_1+L_2+L_3)}_{3n-(L_1+L_2+L_3)}
\]

from which (\ref{5}) follows by splitting the above triple sum into the parts $L_1+L_2+L_3=n $ and $L_1+L_2+L_3\neq n $.
\end{proof}

\begin{remark}\label{rem1}\end{remark} Notice that $3n-(L_1+L_2+L_3)$ is at least equal to $2n$ and 

\[
N(n)^*\,\Phi:=\sum_{L_1=1}^{n-1}\,\sum_{L_2=1}^{n-L_1}\,\sum_{L_3= 1}^{n-(L_1+L_2)}\, b_{L_1}(n,n) \,b_{L_2}(n,n-L_1) 
\]

\[
\times b_{L_3}(n,n-(L_1+L_2)) B_{n-(L_1+L_2+L_3)}^{3n-(L_1+L_2+L_3)}\Phi=0
\]
due to (\ref{dffckvac}) and $n-(L_1+L_2+L_3)\neq 0$.

\begin{remark}\label{rem22}\end{remark} For $n=2$ the previous lemma is not valid since 

\[
C_1(2)=2\,B^0_0+4\,B^1_1,\,\,\,C_2(2)=8\,B^0_2,\,\,\,C_3(2)=0 \Rightarrow \beta (2) =0
\]

Therefore, what follows is not in contradiction with the well established Fock representation of the square of white noise operators $B^2_0$, $B^0_2$ and $B^1_1$ proved in \cite{4}.

\begin{corollary}\label{cor11}Let $n\geq3$ and suppose that an operator $*$-Lie  sub algebra $\mathcal{L}$ of the RPQWN  algebra contains $B^n_0$. Then $\mathcal{L}$ will also contain 

\[
a\,\left(\beta (n)\,B_0^{2n} +N(n)^*\right)+b\,(B^n_0)^2
\]

for all $a,b\in\mathbb{R}$, where $\beta (n)$ and  $N(n)^*$ are as in (\ref{66}) and (\ref{88}) respectively.
\end{corollary}

\begin{proof} Since $\mathcal{L}$ is an operator algebra containing $B^n_0$, it will also contain $(B^n_0)^2$ and $b\,(B^n_0)^2$.  By the $*$-property $\mathcal{L}$  will also contain $B_n^0$ and  since $\mathcal{L}$ is a Lie algebra, by lemma \ref{lem11} , it will contain $\beta (n)\,B^0_{2n} +N(n)$ and $a\,\left(\beta (n)\,B^0_{2n} +N(n)\right)$. Again by the $*$-property,  $\mathcal{L}$ will  contain $a\,\left(\beta (n)\,B_0^{2n} +N(n)^*\right)$ and, since  $\mathcal{L}$ is a vector space, it will also contain $a\,\left(\beta (n)\,B_0^{2n} +N(n)^*\right)+b\,(B^n_0)^2$.
\end{proof}

\begin{theorem}\label{thm22}Let $n\geq3$ and suppose that an operator $*$-Lie  sub algebra $\mathcal{L}$ of the RPQWN  algebra contains $B^n_0$. Then $\mathcal{L}$ does not admit a Fock space representation.
\end{theorem}

\begin{proof} By Corollary \ref{cor11},  $\mathcal{L}$ will also contain $a\,\left(\beta (n)\,B_0^{2n} +N(n)^*\right)+b\,(B^n_0)^2$, for all $a,b\in\mathbb{R}$, where $\beta (n)$, $N(n)^*$ are as in (\ref{66}) and (\ref{88}) respectively. As in the previously discussed counter-example, it follows that the Fock-vacuum norm

\[
\|\left(a\,\left(\beta (n)\,B_0^{2n} +N(n)^*\right)+b\,(B^n_0)^2\right)\Phi\|
=\|\left(a\,B^{2n}_0 +b\,(B^n_0)^2 \right)\Phi\|
\]

cannot be nonnegative for arbitrarily small  $I\subset \mathbb{R}$.
\end{proof}

In the remaining sections of this paper we provide a new renormalization prescription for the powers of the delta function which bypasses the no-go theorems proved so far and which leads to an unexpected connection with the Virasoro algebra and the $w_{\infty}$ and $W_{\infty}$ algebras of Conformal Field Theory (cf. \cite{ketov}).

\section{A new look at the counter-example of the previous section}

In this section we generalize (\ref{f1}) to 

\begin{equation}
\delta^l(t-s)=\phi^{l-1}(s)\,\delta(t-s),\,\,\,\,\,l=2,....\label{f2}
\end{equation}

and we look for conditions on  $\phi (s)$, and an appropriate set of  test functions, that eliminate the difficulties posed by the counter-example of Section 1. The white noise commutation relations (\ref{1}) now become

\begin{equation}
[{b_t^{\dagger}}^nb_t^k,{b_s^{\dagger}}^Nb_s^K]=\label{3}
\end{equation}

\[
\epsilon_{k,0}\epsilon_{N,0}\sum_{L\geq 1} \binom{k}{L}N^{(L)}\,{b_t^{\dagger}}^{n}\,{b_s^{\dagger}}^{N-L}\,b_t^{k-L}\,b_s^
K\,\phi^{L-1}(s)\,\delta(t-s)
\]

\[
-\epsilon_{K,0}\epsilon_{n,0}\sum_{L\geq 1} \binom{K}{L}n^{(L)}\,{b_s^{\dagger}}^{N}\,{b_t^{\dagger}}^{n-L}\,b_s^{K-L}\,b_t^k\,\phi^{L-1}(s)\,\delta(t-s)
\]

from which,  by multiplying both sides  by  $f(t)\bar g (s)$ and  integrating the resulting identity we obtain

\begin{equation}
[B^N_K(\bar g),B^n_k(f) ]=\label{4}
\end{equation}

\[
\sum_{L= 1}^{K \wedge n}\, \hat{b}_L(K,n)\,B^{N+n-L}_{K+k-L}(\bar g\, f\,\phi^{\,L-1})-\sum_{L= 1}^{k \wedge N }\,\hat{b}_L(k,N)\,B^{N+n-L}_{K+k-L}(\bar g\, f\,\phi^{L-1})
\]

\[
=\sum_{L= 1}^{ (K \wedge n) \vee ( k \wedge N)  }\, \theta_L (N,K;n,k)
\,B^{N + n - L}_{K + k - L} (\bar g \,f \,\phi^{L-1})
\]

where

\[
\hat{b}_x(y,z):=\epsilon_{y,0}\,\epsilon_{z,0}\, \binom{y}{x}\,z^{(x)}
\]

$n,k,N,K\in \{0,1,2,...\}$, and $\theta_L(N,K;n,k)$ is as in (\ref{2a}).  Turning to the counter-example of Section 1, for an interval  $I \subset \mathbb{R}$, introducing the notation

\[
I_n=\int_I\,\phi^n(s)\,ds,\,\,\,\,\,n=0,1,2,...
\]

and using commutation relations (\ref{4}) we have for $n\geq 1$

\[
B^0_{2n}(\chi_I)\,B^{2n}_0(\chi_I)\,\Phi=[B^0_{2n}(\chi_I),B^{2n}_0(\chi_I)]\,\Phi
\]

\[
=\sum_{L=1}^{2n}\,\binom{2n}{L}(2n)^{(L)}\,B^{2n-L}_{2n-L}(\phi^{L-1}\,\chi_I)\,\Phi
\]

\[
=\binom{2n}{2n}(2n)^{(2n)}\,B^{0}_{0}(\phi^{2n-1}\,\chi_I)\,\Phi\\
=(2n)!\,\int_I\,\phi^{2n-1}(s)\,ds\,\Phi
\]

and so

\[
<B^0_{2n}(\chi_I)\,B^{2n}_0(\chi_I)>=(2n)!\,\int_I\,\phi^{2n-1}(s)\,ds=(2n)!\,I_{2n-1}
\]

Similarly,

\[
B^0_{2n}(\chi_I)\,(B^{n}_0(\chi_I))^2\,\Phi=
\]

\[
\left(B^n_0(\chi_I)\,B^0_{2n}(\chi_I)+[B^0_{2n}(\chi_I),B^{n}_0(\chi_I)]\right)\,B^n_0(\chi_I)\,\Phi
\]

\[
=B^n_0(\chi_I)\,B^0_{2n}(\chi_I)\,B^n_0(\chi_I)\,\Phi+[B^0_{2n}(\chi_I),B^{n}_0(\chi_I)]\,B^n_0(\chi_I)\,\Phi
\]

\[
=B^n_0(\chi_I)\,[B^0_{2n}(\chi_I),B^n_0(\chi_I)]\,\Phi+[B^0_{2n}(\chi_I),B^{n}_0(\chi_I)]\,B^n_0(\chi_I)\,\Phi
\]

\[
=B^n_0(\chi_I)\,\sum_{L=1}^n\,\hat{b}_L(2n,n)\,B^{n-L}_{2n-L}(\phi^{L-1}\,\chi_I)\,\Phi 
\]

\[
+\sum_{L=1}^n\,\hat{b}_L(2n,n)\,B^{n-L}_{2n-L}(\phi^{L-1}\,\chi_I)\,B^n_0(\chi_I)\,\Phi
\]

\[
=0+\sum_{L=1}^n\,\hat{b}_L(2n,n)\,[B^{n-L}_{2n-L}(\phi^{L-1}\,\chi_I),B^n_0(\chi_I)]\,\Phi
\]

\[
=\sum_{L_1=1}^n\,\sum_{L_2=1}^n\,\hat{b}_{L_1}(2n,n)\,\hat{b}_{L_2}(2n-L_1,n)\,B^{2n-(L_1+L_2)}_{2n-(L_1+L_2)}(\phi^{L_1+L_2-2}\,\chi_I)\,\Phi
\]

\[
=\hat{b}_{n}(2n,n)\,\hat{b}_{n}(n,n)\,B^0_0(\phi^{2n-2}\,\chi_I)\,\Phi\\
=(2n)!\,\int_I\,\phi^{2n-2}(s)\,ds\,\Phi
\]

which implies that

\[
<B^0_{2n}(\chi_I)\,(B^{n}_0(\chi_I))^2>=(2n)!\,\int_I\,\phi^{2n-2}(s)\,ds=(2n)!\,I_{2n-2}
\]

We also have

\[
B^0_{n}(\chi_I)\,(B^{n}_0(\chi_I))^2\,\Phi=
\]

\[
\left(B^{n}_0(\chi_I)\,B^0_{n}(\chi_I)+[ B^0_{n}(\chi_I),B^{n}_0(\chi_I) ]\right)\,B^{n}_0(\chi_I)\,\Phi
\]

\[
=B^{n}_0(\chi_I)\,B^0_{n}(\chi_I)\,B_0^{n}(\chi_I)\,\Phi+[ B^0_{n}(\chi_I),B^{n}_0(\chi_I) ]\,B^{n}_0(\chi_I)\,\Phi
\]

\[
=B^{n}_0(\chi_I)\,\left(B^{n}_0(\chi_I)\,B^0_{n}(\chi_I)+[ B^0_{n}(\chi_I),B^{n}_0(\chi_I) ]\right)\,\Phi+[B^0_{n}(\chi_I),B^{n}_0(\chi_I)]\,B^{n}_0(\chi_I)\,\Phi
\]

\[
=B^{n}_0(\chi_I)\,[ B^0_{n}(\chi_I),B^{n}_0(\chi_I) ]\,\Phi+[B^0_{n}(\chi_I),B^{n}_0(\chi_I)]\,B^{n}_0(\chi_I)\,\Phi
\]

\[
=B^{n}_0(\chi_I)\,\sum_{L=1}^{n}\, \hat{b}_L(n,n) \,B^{n-L}_{n-L}(\phi^{L-1}\,\chi_I)\,\Phi
+\sum_{L=1}^{n}\, \hat{b}_L(n,n) \,B^{n-L}_{n-L}(\phi^{L-1}\,\chi_I)\,B^{n}_0(\chi_I)\,\Phi
\]

\[
=B^{n}_0(\chi_I)\, \hat{b}_n(n,n) \,B^0_0(\phi^{n-1}\,\chi_I)\,\Phi\
\]

\[
+\sum_{L=1}^{n}\, \hat{b}_L(n,n) \,\left(B^{n}_0(\chi_I)\, B^{n-L}_{n-L}(\phi^{L-1}\,\chi_I)+[ B^{n-L}_{n-L}(\phi^{L-1}\,\chi_I),B^{n}_0(\chi_I)]\right)\Phi
\]

\[
=\hat{b}_n(n,n) \,\int_I\,\phi^{n-1}(s)\,ds\,B^{n}_0(\chi_I)\,\Phi+\hat{b}_n(n,n) \,B^{n}_0(\chi_I)\,B^0_0(\phi^{n-1}\,\chi_I)\,\Phi
\]

\[
+\sum_{L_1=1}^n\,\sum_{L_2=1}^{n-L_1}\,\hat{b}_{L_1}(n,n) \,\hat{b}_{L_2}(n-L_1,n) \,B^{2n-(L_1+L_2)}_{n-(L_1+L_2)}(\phi^{L_1+L_2-2}\,\chi_I)\,\Phi
\]

\[
=2\,\hat{b}_n(n,n) \,\int_I\,\phi^{n-1}(s)\,ds\,B^{n}_0(\chi_I)\,\Phi+\sum_{L=1}^{n-1}\,\hat{b}_{L}(n,n) \,\hat{b}_{n-L}(n-L,n) \,B^n_0(\phi^{n-2}\,\chi_I)\,\Phi
\]

\[
=2\,(n!) \,\int_I\,\phi^{n-1}(s)\,ds\,B^{n}_0(\chi_I)\,\Phi+\left((2n)^{(n)}-2\,(n!)\right)\,B^n_0(\phi^{n-2}\,\chi_I)\,\Phi
\]

Thus

\[
(B^0_{n}(\chi_I))^2\,(B^{n}_0(\chi_I))^2\,\Phi=
\]

\[
2\,(n!) \,I_{n-1}(s)\,B_{n}^0(\chi_I)\,B^{n}_0(\chi_I)\,\Phi+\left((2n)^{(n)}-2\,(n!)\right)\,B_{n}^0(\chi_I)\,B^n_0(\phi^{n-2}\,\chi_I)\,\Phi
\]

\[
=2\,(n!) \,I_{n-1}\,[B_{n}^0(\chi_I),B^{n}_0(\chi_I)]\,\Phi+\left((2n)^{(n)}-2\,(n!)\right)\,[B_{n}^0(\chi_I),B^n_0(\phi^{n-2}\,\chi_I)]\,\Phi
\]

\[
=2\,(n!)^2 \,(I_{n-1})^2\,\Phi+\left((2n)^{(n)}-2\,(n!)\right)\,\sum_{L=1}^{n}\,\hat{b}_{L}(n,n)\,B^{n-L}_{n-L}(\phi^{n-2+L-1}\,\chi_I)\,\Phi
\]

\[
=2\,(n!)^2 \,(I_{n-1})^2\,\Phi+\left((2n)^{(n)}-2\,(n!)\right)\,\hat{b}_{n}(n,n)\,B^{0}_{0}(\phi^{2n-3}\,\chi_I)\,\Phi
\]

\[
=2\,(n!)^2 \,(I_{n-1})^2\,\Phi+\left((2n)^{(n)}-2\,(n!)\right)\,(n!)\,I_{2n-3}\,\Phi
\]

\[
=2\,(n!)^2 \,(I_{n-1})^2\,\Phi+\left((2n)!-2\,(n!)^2\right)\,I_{2n-3}\,\Phi
\]

and so

\[
<(B^0_{n}(\chi_I))^2\,(B^{n}_0(\chi_I))^2>=2\,(n!)^2 \,(I_{n-1})^2+\left((2n)!-2\,(n!)^2\right)\,I_{2n-3}
\]

Thus the matrix $A$ of the counter-example of Section 1 has the form

\[
A=\left[
\begin{array}{ccc}
 <B^0_{2n}(\chi_I)\,B^{2n}_0(\chi_I)>&& <B^0_{2n}(\chi_I)\,(B^n_0(\chi_I))^2>\\
<B^0_{2n}(\chi_I)\,(B^n_0(\chi_I))^2> &&< (B^0_n(\chi_I))^2\,(B^n_0(\chi_I))^2>
\end{array}
\right]
\]

\[
=\left[
\begin{array}{ccc}
     (2n)!\,I_{2n-1}  & & (2n)!\,I_{2n-2}   \\
        (2n)!\,I_{2n-2} & &2\,(n!)^2 \,(I_{n-1})^2+\left((2n)!-2\,(n!)^2\right)\,I_{2n-3}
\end{array}
\right]
\]

with minor determinants

\[
d_1=(2n)!\,I_{2n-1} 
\]

which will be $\geq0$ if  

\begin{equation}
I_{2n-1}\geq0\label{c1}
\end{equation}

for all $n$ and  $I \subset \mathbb{R}$,  and

\[
d_2= (2n)!\,(2\,(n!)^2 \,I_{2n-1} \, (I_{n-1})^2
\]

\[
+\,\,\,((2n)!-2\,(n!)^2)\,I_{2n-1}\,I_{2n-3}
-(2n)!\,(I_{2n-2})^2)
\]

which will be $\geq0$ if 

\[
2\,(n!)^2 \,I_{2n-1} \, (I_{n-1})^2+\left((2n)!-2\,(n!)^2\right)\,I_{2n-1}\,I_{2n-3}
-(2n)!\,(I_{2n-2})^2\geq0
\]

i.e if

\[
\left((2n)!-2\,(n!)^2\right)\,I_{2n-1}\,I_{2n-3}\geq (2n)!\,(I_{2n-2})^2 -2\,(n!)^2 \,I_{2n-1} \, (I_{n-1})^2
\]

which will be satisfied if

\begin{equation}
(I_{2n-2})^2=I_{2n-1}\,I_{2n-3}  \label{c2}
\end{equation}

and

\begin{equation}
 I_{2n-1} \, (I_{n-1})^2\geq I_{2n-1}\,I_{2n-3}\label{c3}
\end{equation}

for all $n$ and  $I \subset \mathbb{R}$. It was condition (\ref{c3}) that created all the trouble in the counter-example of Section 1.

\section{ A renormalization suggested by conditions (\ref{c1})-(\ref{c3})}

We notice that if $\mbox{supp} (\phi) \cap I=\emptyset$ then conditions (\ref{c1})-(\ref{c3}) are trivially satisfied. If $ \mbox{supp} (\phi)  \cap I\neq\emptyset$ then conditions (\ref{c1})-(\ref{c3}) are  satisfied by $I_n=1$ for all $n=1,2,...$, which is true if $\phi^n=\delta$ for all $n=1,2,...$.  The renormalization rule (\ref{f2}) then becomes

\begin{equation}
\delta^l(t-s)=\delta(s)\,\delta(t-s),\,\,\,\,\,l=2,3,....\label{f3}
\end{equation}

and (\ref{1}) takes the form

\begin{equation}
[{b_t^{\dagger}}^nb_t^k,{b_s^{\dagger}}^Nb_s^K]=\label{5}
\end{equation}

\[
\epsilon_{k,0}\epsilon_{N,0}\,( k\,N\, {b_t^{\dagger}}^{n}\,{b_s^{\dagger}}^{N-1}\,b_t^{k-1}\,b_s^k\,\delta(t-s)
\]

\[
 +    \sum_{L\geq 2} \binom{k}{L}N^{(L)}\,{b_t^{\dagger}}^{n}\,{b_s^{\dagger}}^{N-L}\,b_t^{k-L}\,b_s^K\, \delta(s)\,\delta(t-s) )
\]

\[
-\epsilon_{K,0}\epsilon_{n,0}( K\,n \,{b_s^{\dagger}}^{N}\,{b_t^{\dagger}}^{n-1}\,b_s^{K-1}\,b_t^k\,\delta(t-s) 
\]

\[
  +   \sum_{L\geq 2} \binom{K}{L}n^{(L)}\,{b_s^{\dagger}}^{N}\,{b_t^{\dagger}}^{n-L}\,b_s^{K-L}\,b_t^k\,\delta(s)\,\delta(t-s))
\]

which, after  multiplying both sides  by  $f(t)\bar g (s)$ and  integrating the resulting identity, yields the commutation relations

\begin{equation}
[B^n_k(\bar g),B^N_K(f) ]=\left(\epsilon_{k,0}\epsilon_{N,0}\, k\,N- \epsilon_{K,0}\epsilon_{n,0}\, K\,n     \right)\, B^{N+n-1}_{K+k-1}(\bar g f) \label{6}
\end{equation}

\[
+\sum_{L= 2}^{ (K \wedge n) \vee ( k \wedge N)  }\, \theta_L (n,k;N,K)\,\bar g(0) \,f(0)\,{b_0^{\dagger}}^{N+n-l}\,b_0^{K+k-l}  
\]

where  $\theta_L(n,k;N,K)$ is as in (\ref{2a}).  We can write (\ref{6}) as

\begin{equation}
[B^n_k(\bar g),B^N_K(f) ]=\left(\epsilon_{k,0}\epsilon_{N,0}\, k\,N- \epsilon_{K,0}\epsilon_{n,0}\, K\,n     \right)\, B^{N+n-1}_{K+k-1}(\bar g f)\label{7}
\end{equation}

\[
+\sum_{L= 2}^{ (K \wedge n) \vee ( k \wedge N)  }\, \theta_L (n,k;N,K)\,   B^{N+n-L}_{K+k-L}(\bar g \, f\, \delta)       
\]

and notice that repeated commutations with the use of (\ref{7}) will introduce terms containing $\delta(0)$. 

\section{The canonical RPQWN commutation relations}

We may eliminate the singular terms from (\ref{6}) by restricting to test functions $f$ that satisfy $f(0)=0$. We then define the canonical RPQWN commutation relations as follows.

\begin{definition}\label{rpqwn} For right-continuous step functions $f,g$ such that $f(0)=g(0)=0$ we define

\begin{equation}
[B^n_k(\bar g),B^N_K(f) ]_{R}:= \left( k\,N- K\,n     \right)\, B^{n+N-1}_{k+K-1}(\bar g f)\label{8} 
\end{equation}

\end{definition}

Letting

\begin{equation}\label{mtx}
C(n,k;N,K):=\left[
\begin{array}{ccc}
 N&&n \\
   K &&k
\end{array}
\right]
\end{equation}

commutation relations (\ref{8}) can also be written as

\begin{equation}
[B^n_k(\bar g),B^N_K(f) ]_{R}= \det C(n,k;N,K) \, B^{n+N-1}_{k+K-1}(\bar g f)\label{9} 
\end{equation} 

\begin{proposition}

Commutation relations (\ref{8}) define a Lie algebra.

\end{proposition}

\begin{proof} Clearly 

\[
[B^N_K(\bar g),B^N_K(f) ]_{R}=0
\]

and

\[
[B^N_K(\bar g),B^n_k(f) ]_{R}=-[B^n_k(f),B^N_K(\bar g) ]_{R}
\]

To show that commutation relations (\ref{8}) satisfy the Jacobi identity we must show that (suppressing the test functions $f$ and $g$) for all $n_i,k_i\geq0$, where  $i=1,2,3$,

\[
\left[B^{n_1}_{k_1},[B^{n_2}_{k_2},B^{n_3}_{k_3}]_{R}\right]_{R}+ \left[B^{n_3}_{k_3},[B^{n_1}_{k_1},B^{n_2}_{k_2}]_{R}\right]_{R}+\left[B^{n_2}_{k_2},[B^{n_3}_{k_3},B^{n_1}_{k_1}]_{R}\right]_{R}  =0
\]

i.e.  that 

\[
\det C(n_2,k_2;n_3,k_3)\,\det C(n_1,k_1;n_2+n_3-1,k_2+k_3-1)+
\]

\[
\det C(n_1,k_1;n_2,k_2)\,\det C(n_3,k_3;n_1+n_2-1,k_1+k_2-1)+
\]

\[
\det C(n_3,k_3;n_1,k_1)\,\det C(n_2,k_2;n_3+n_1-1,k_3+k_1-1)=0
\]

which is the same as

\[
(n_2 k_3 - n_3 k_2)(n_1 k_2 + n_1 k_3 - n_1 - k_1 n_2 - k_1 n_3 + k_1)
  +
\]

\[
 (n_1 k_2 - n_2 k_1)(n_3 k_1 + n_3 k_2 - n_3 - n_1 k_3 - n_2 k_3 + k_3)
  +
\]

\[
 (n_3 k_1 - n_1 k_3)(n_2 k_3 + n_2 k_1 - n_2 - n_3 k_2 - n_1 k_2 + k_2)=0
\]

and is easily seen to be true.

\end{proof}

\section{The  $w_{\infty}$ algebra}

\begin{definition} The $w_{\infty}$ algebra (see  \cite{4a}, \cite{ketov})  is the infinite dimensional non-associative Lie algebra spanned by the generators $\hat{B}^n_k$, where $n,k\in\mathbb{Z}$ with $n\geq 2$,  with commutation relations

\begin{equation}
[ \hat{B}^n_k , \hat{B}^N_K ]_{w_{\infty} } =\left(   (N-1)\,k-(n-1)\,K  \right) \,\hat{B}^{n+N-2}_{k+K}\label{vir} 
\end{equation}

and adjoint condition

\begin{equation}
\left( \hat{B}^n_k \right)^*= \hat{B}^n_{-k} \label{adjvir} 
\end{equation}

\end{definition}

The $w_{\infty}$ algebra is the basic algebraic structure  of Conformal Field Theory in the study of quantum membranes. Since it contains as a sub algebra the Virasoro  algebra with commutations

\[
[\hat{B}^2_k(\bar g),\hat{B}^2_{K}(f) ]_{V}:=(k-K)\,\hat{B}^{2}_{k+k}(\bar g f)
\]

$w_{\infty}$ can be viewed as an extended conformal algebra with an infinite number of additional symmetries (see \cite{4a}-\cite{4d}, \cite{ketov},  \cite{7}, \cite{8}).    The elements of $w_{\infty}$ are interpreted as area preserving diffeomorphisms of 2-manifolds.  A  quantum deformation of  $w_{\infty}$, called $W_{\infty}$ and defined as a, large $N$, limit of Zamolodchikov's $W_{N}$ algebra (see \cite{8}), has been studied extensively  ( see \cite{4b}-\cite{4d}, \cite{ketov}, \cite{7}) in connection to two-dimensional Conformal Field Theory and Quantum Gravity.  $w_{\infty}$  is a "classical" or "Gel'fand-Dikii" algebra (see \cite{GD}) in the sense that it is a $W$ algebra (see \cite{ketov}) where all central terms are set to zero.

\section{Poisson brackets}

The construction produced in the following section was inspired by the analogy with 
the realization of the $w$--algebra in terms of Poisson brackets. This realization is well
known and, in the following, we recall it briefly.

\begin{definition} For scalar-valued differentiable functions $f(x,y)$ and $g(x,y)$,  the Poisson bracket $\{f,g\}$ is defined by

\[
\{f,g\}=\frac{ \partial f }{  \partial x}\,\frac{ \partial g }{ \partial  y}-\frac{ \partial f }{ \partial y}\, \frac{ \partial g }{  \partial x }
\]

\end{definition}

We notice that the functions $f(x,y)=x$ and $g(x,y)=y$ satisfy $\{f,g\}=1$ which we can write as

\[
\{x,y\}=1
\]

in analogy with the Canonical Commutation Relations (CCR).  We can model commutation relations (\ref{vir}) and the adjoint condition (\ref{adjvir}) using the Poisson bracket as follows:

\begin{proposition}\label{pp1} For $n,k \in \mathbb{Z}$ with $n\geq 2$, let $f_{n,k}:\mathbb{R}\times\mathbb{R}\rightarrow\mathbb{C}$  be defined by $f_{n,k}(x,y)=e^{ikx}\,y^{n-1}$. Then

\begin{equation}
\{f_{n,k}(x,y),f_{N,K}(x,y)\}=i\,\left(k(N-1)-K(n-1) \right) \,f_{n+N-2,k+K}(x,y)\label{poi}
\end{equation}

and

\[
\overline{f}_{n,k}(x,y)=f_{n,-k}(x,y)
\]

\end{proposition}

\begin{proof} By the  definition of the Poisson bracket, 

\[
\{f_{n,k}(x,y),f_{N,K}(x,y)\}=\frac{ \partial  }{  \partial x}( e^{ikx}\,y^{n-1}   )\,\frac{ \partial  }{ \partial  y}( e^{iKx}\,y^{N-1})
\]

\[
-\frac{ \partial  }{ \partial y }( e^{ikx}\,y^{n-1}    )\, \frac{ \partial  }{  \partial x }( e^{iKx}\,y^{N-1})
\]

\[
= i\,\left( k(N-1)  -K(n-1)\right)\, e^{i(k+K)x}\,y^{n+N-3}
\]

\[
=i\,\left(k(N-1)  -K(n-1)  \right)\, f_{n+N-2,k+K}(x,y)
\]

Moreover,

\[
\overline{f}_{n,k}(x,y)=\overline{e^{ikx}\,y^{n-1}  }= e^{-ikx}\,y^{n-1}  =f_{n,-k}(x,y)
\]

\end{proof}

Using the prescription 

\[
[A,B]=\frac{\hbar}{i}\,\{A,B\}
\]

we thus obtain, letting $\hbar=1$, that the quantized version of  (\ref{poi}) is

\[
[f_{n,k},f_{N,K}]=\left( k(N-1) - K(n-1)   \right) \,f_{n+N-2,k+K}
\]

which is precisely (\ref{vir}). Similarly, we can model commutation relations (\ref{8}) and the RPQWN adjoint condition $\left( B^n_k \right)^*= B^k_n$  using the Poisson bracket as follows:

\begin{proposition}\label{pp2} For $n,k\geq0$, let  $g_{n,k}:\mathbb{R}\times \mathbb{R}\rightarrow \mathbb{C}$ be defined by  

\[
g_{n,k}(x,y)=\left(\frac{x+iy}{\sqrt{2}}\right)^n\left(\frac{x-iy}{ \sqrt{2}}\right)^k
\]

Then

\begin{equation}
\{g_{n,k}(x,y),g_{N,K}(x,y)\}=i\,\left( kN-nK \right) \,g_{n+N-1,k+K-1}(x,y) \label{rpoi}
\end{equation}

and

\[
\overline{g}_{n,k}(x,y)=g_{k,n}(x,y)
\]

\end{proposition}

\begin{proof} By the  definition of the Poisson bracket, 

\[
\{g_{n,k}(x,y),g_{N,K}(x,y)\}=
\]

\[
\frac{ \partial}{\partial x}\left(  \left(\frac{x+iy}{\sqrt{2}}\right)^n\left(\frac{x-iy}{ \sqrt{2}}\right)^k   \right)\,\frac{ \partial  }{ \partial  y} 
\left( \left(\frac{x+iy}{\sqrt{2}}\right)^N\left(\frac{x-iy}{ \sqrt{2}}\right)^K  \right)
\]

\[
-\frac{ \partial  }{ \partial y }\left( \left(\frac{x+iy}{\sqrt{2}}\right)^n\left(\frac{x-iy}{ \sqrt{2}}\right)^k    \right)\, \frac{ \partial  }{  \partial x } \left(  \left(\frac{x+iy}{\sqrt{2}}\right)^N\left(\frac{x-iy}{ \sqrt{2}}\right)^K      \right)
\]

\[
= i\,\left( kN-nK \right)\,2^{1-\frac{ n+k+N+K  }{2 }}
\,( x+iy)^{n+N-1} ( x-iy)^{k+K-1} 
\]

\[
= i\,\left( kN-nK \right)
\,\left( \frac{x+iy}{\sqrt{2}} \right )^{n+N-1}  \left( \frac{x-iy}{ \sqrt{2}} \right)^{k+K-1}
\]

\[
= i\,\left( kN-nK \right)\, g_{n+N-1,k+K-1}(x,y)
\]

Moreover,

\[
\overline{g}_{n,k}(x,y)=\overline{ \left(\frac{x+iy}{\sqrt{2}}\right)^n\left(\frac{x-iy}{ \sqrt{2}}\right)^k  }= \left(\frac{x-iy}{\sqrt{2}}\right)^n\left(\frac{x+iy}{ \sqrt{2}}\right)^k =g_{k,n}(x,y)
\]

\end{proof}

We therefore have, as above, that the quantized version of  (\ref{rpoi}) is

\[
[g_{n,k},g_{N,K}]=\left( kN-nK    \right) \,g_{n+N-1,k+K-1}
\]

which is  (\ref{8}).

\section{White noise form of the $w_{\infty}$ generators and commutation relations}

Motivated by the results of the previous section we introduce the following:

\begin{definition}\label{B-hat}
For right-continuous step functions $f,g$ such that $f(0)=g(0)=0$, and for $n,k\in\mathbb{Z}$  with $n\geq2$, we define

\begin{equation}
\hat{B}_k^n(f):=\int_{\mathbb{R}}\,f(t)\,e^{ \frac{k}{2}(b_t- b_t^{\dagger})}\left(\frac{ b_t+ b_t^{\dagger}}{2}\right)^{n-1} \,  e^{ \frac{k}{2}(b_t- b_t^{\dagger})}\,dt\label{op}
\end{equation}

with involution

\[
\left(\hat{B}_k^n(f)\right)^{*} = \hat{B}_{-k}^n(\bar f)
\]

In particular,

\begin{equation}
\hat{B}_k^2(f):=\int_{\mathbb{R}}\,f(t)\,e^{ \frac{k}{2}(b_t- b_t^{\dagger})}\left(\frac{ b_t+ b_t^{\dagger}}{2}\right) \,  e^{ \frac{k}{2}(b_t- b_t^{\dagger})}\,dt\label{vira}
\end{equation}

is the RPQWN form of the Virasoro operators

The integral on the right hand side of (\ref{op}) is meant in the sense that one expands 
the exponential series (resp. the power), applies the commutation relations (\ref{1}) to bring 
the resulting expression to normal order, introduces the renormalization prescription 
(\ref{f3}), integrates the resulting expressions after multiplication by a test function 
and interprets the result  as a quadratic form on the exponential vectors.
\end{definition}

\begin{lemma}\label{f} 
Let $x$ , $ D$  and $h$ be three operators satisfying the Heisenberg commutation relations

\[
[D,x]=h, \,\,\,[D,h]=[x,h]=0
\]

Then, for all $s,a,c\in\mathbb{C}$

\[
e^{s(x+aD+ch)}=e^{sx}e^{saD}e^{(sc+\frac{s^2a}{2})h}
\]

\[
e^{sD}e^{ax}=e^{ax}e^{sD}e^{ash}
\]

and for all $n,m\in\mathbb{N}$

\[
D^nx^m=\sum_{j=1}^{n\wedge m}\binom{n,m}{j}x^{m-j}D^{n-j}h^j
\]

where

\[
\binom{n,m}{j}=\binom{n}{j}\binom{m}{j}j!
\]

\end{lemma}

\begin{proof}
This is just a combination of Propositions 2.2.2, 2.2.1 and 4.1.1 of \cite{fein}.
\end{proof}

\begin{lemma}\label{ff} In the notation of lemma \ref{f}, for all $\lambda \in \{0,1,...\}$ and  
$a\in\mathbb{C}$

\[
D^{\lambda}e^{ax}=e^{ax}\sum_{m=0} ^{\lambda}\binom{ \lambda }{ m }D^m (ah)^{\lambda-m}
\]

and

\[
e^{sD}x^{\lambda}=\sum_{m=0} ^{\lambda}\binom{ \lambda }{ m }x^m (sh)^{\lambda-m}e^{sD}
\]

\end{lemma}

\begin{proof} By lemma \ref{f}

\[
D^{\lambda }e^{ax}=\frac{ \partial^{\lambda}}{\partial s^{\lambda}}|_{s=0}\left(e^{sD}e^{ax}\right)=
\frac{ \partial^{\lambda} }{ \partial s^{\lambda} }|_{s=0}\left( e^{ax}e^{sD}e^{ash}  \right)
=e^{ax} \frac{ \partial^{\lambda}}{ \partial s^{\lambda} }
|_{s=0}\left( e^{sD}e^{ash}  \right)
\]

\[
=e^{ax}\sum_{m=0} ^{\lambda}\binom{ \lambda }{ m }
\frac{ \partial^m }{\partial s^m}|_{s=0}    \left( e^{sD} \right)\,\frac{\partial^{\lambda-m}}{\partial s^{\lambda -m}}|_{s=0}\left(e^{ash} \right)
=e^{ax}\sum_{m=0} ^{\lambda}\binom{ \lambda }{ m }D^m (ah)^{\lambda-m}
\]

Similarly,

\[
e^{sD}x^{\lambda}=\frac{ \partial^{\lambda}}{\partial a^{\lambda}}|_{a=0}\left(e^{sD}e^{ax}\right)=
\frac{ \partial^{\lambda} }{ \partial a^{\lambda} }|_{a=0}\left( e^{ax}e^{sD}e^{ash}  \right)
= \frac{ \partial^{\lambda}}{ \partial a^{\lambda} }
|_{a=0}\left(e^{ax} e^{ash}  \right)e^{sD}
\]

\[
=\sum_{m=0} ^{\lambda}\binom{ \lambda }{ m }
\frac{ \partial^m }{\partial a^m}|_{a=0}    \left(e^{ax}  \right)\,\frac{\partial^{\lambda-m}}{\partial a^{\lambda -m}}|_{a=0}\left(e^{ash} \right)e^{sD}
=\sum_{m=0} ^{\lambda}\binom{ \lambda }{ m }x^m (sh)^{\lambda-m}e^{sD}
\]

\end{proof}

\begin{lemma}\label{fff} Let the exponential and powers of white noise be interpreted as 
described in Definition (\ref{B-hat}). Then:

(i) For fixed $t,s\in\mathbb{R}$, the operators $D= b_t- b_t^{\dagger}$, 
$x= b_s+ b_s^{\dagger}$ and $h=2\,\delta(t-s)$ satisfy the commutation relations of lemma \ref{f}.

(ii) For fixed $t,s\in\mathbb{R}$, the operators $D= b_t+ b_t^{\dagger}$, $x= b_s- b_s^{\dagger}$ and $h=-2\,\delta(t-s)$ satisfy the commutation relations of lemma \ref{f}.
\end{lemma}

\begin{proof} To prove (i) we notice that

\[
[D,x]=[b_t- b_t^{\dagger} ,b_s+ b_s^{\dagger} ]= [b_t ,b_s^{\dagger} ]-[ b_t^{\dagger} ,b_s]= [b_t ,b_s^{\dagger} ]+[b_s, b_t^{\dagger} ]=\delta(t-s)+\delta(s-t)=h 
\]

while, clearly, $[D,h]=[x,h]=0$. The proof of (ii) is similar.

\end{proof}

\begin{proposition}\label{comw} If  $f,g$ are right-continuous step functions such that $f(0)=g(0)=0$ and the powers of the delta function are renormalized by the prescription (\ref{f3}), then

\begin{equation}
 [\hat{B}^n_k(\bar g),\hat{B}^N_K(f) ]= \left( k\,(N-1)- K\,(n-1) \right)\, \hat{B}^{n+N-2}_{k+K}(\bar g f)\label{wcom} 
\end{equation}

i.e the operators $\hat{B}_k^n$ of Definition \ref{B-hat} satisfy the commutation relations of the $w_{\infty}$ algebra. In particular, 

\begin{equation}
[\hat{B}^2_k(\bar g),\hat{B}^2_K(f) ]= \left( k- K \right)\, \hat{B}^{2}_{k+K}(\bar g f)\label{viras}
\end{equation}

i.e the operators $\hat{B}_k^2$ of Definition \ref{B-hat} satisfy the commutation relations of the Virasoro algebra.
Here $[x,y]:=xy-yx$ is the usual operator commutator.

\end{proposition}

\begin{proof} To prove (\ref{wcom}), we notice that by Definition \ref{B-hat}, its left hand side is

\[
\int_{\mathbb{R}}\int_{\mathbb{R}}\bar{g}(t)f(s) [e^{ \frac{k}{2}(b_t- b_t^{\dagger})}\left(\frac{ b_t+ b_t^{\dagger}}{2}\right)^{n-1}   e^{ \frac{k}{2}(b_t- b_t^{\dagger})},
\]

\[
 e^{ \frac{K}{2}(b_s- b_s^{\dagger})}\left(\frac{ b_s+ b_s^{\dagger}}{2}\right)^{N-1}   e^{ \frac{K}{2}(b_s- b_s^{\dagger})}] \,dt\,ds
\]

\[
=\int_{\mathbb{R}}\int_{\mathbb{R}}\bar{g}(t)f(s)\, e^{ \frac{k}{2}(b_t- b_t^{\dagger})}\left(\frac{ b_t+ b_t^{\dagger}}{2}\right)^{n-1}   e^{ \frac{k}{2}(b_t- b_t^{\dagger})} 
\]

\[
\times
 e^{ \frac{K}{2}(b_s- b_s^{\dagger})}\left(\frac{ b_s+ b_s^{\dagger}}{2}\right)^{N-1}   e^{ \frac{K}{2}(b_s- b_s^{\dagger})}    \,dt\,ds
\]

\[
-\int_{\mathbb{R}}\int_{\mathbb{R}}\bar{g}(t)f(s)\, e^{ \frac{K}{2}(b_s- b_s^{\dagger})}\left(\frac{ b_s+ b_s^{\dagger}}{2}\right)^{N-1}   e^{ \frac{K}{2}(b_s- b_s^{\dagger})} 
\]

\[
\times
 e^{ \frac{k}{2}(b_t- b_t^{\dagger})}\left(\frac{ b_t+ b_t^{\dagger}}{2}\right)^{n-1}   e^{ \frac{k}{2}(b_t- b_t^{\dagger})}   \,dt\,ds
\]

which, since $[b_t- b_t^{\dagger} ,b_s+ b_s^{\dagger} ]=0$, is

\[
=\int_{\mathbb{R}}\int_{\mathbb{R}}\bar{g}(t)f(s)\, e^{ \frac{k}{2}(b_t- b_t^{\dagger})}\left(\frac{ b_t+ b_t^{\dagger}}{2}\right)^{n-1}    e^{ \frac{K}{2}(b_s- b_s^{\dagger})}
\]

\[
\times  e^{ \frac{k}{2}(b_t- b_t^{\dagger})}\left(\frac{ b_s+ b_s^{\dagger}}{2}\right)^{N-1}   e^{ \frac{K}{2}(b_s- b_s^{\dagger})}    \,dt\,ds
\]

\[
-\int_{\mathbb{R}}\int_{\mathbb{R}}\bar{g}(t)f(s)\, e^{ \frac{K}{2}(b_s- b_s^{\dagger})}\left(\frac{ b_s+ b_s^{\dagger}}{2}\right)^{N-1}    e^{ \frac{k}{2}(b_t- b_t^{\dagger})}
\]

\[
\times e^{ \frac{K}{2}(b_s- b_s^{\dagger})} \left(\frac{ b_t+ b_t^{\dagger}}{2}\right)^{n-1}   e^{ \frac{k}{2}(b_t- b_t^{\dagger})}   \,dt\,ds
\]

\[
=\frac{ 1  }{ 2^{n+N-2}}\{\int_{\mathbb{R}}\int_{\mathbb{R}}\bar{g}(t)f(s)\, e^{ \frac{k}{2}(b_t- b_t^{\dagger})}(b_t+ b_t^{\dagger})^{n-1}    e^{ \frac{K}{2}(b_s- b_s^{\dagger})} 
\]

\[
\times e^{ \frac{k}{2}(b_t- b_t^{\dagger})}( b_s+ b_s^{\dagger})^{N-1}   e^{ \frac{K}{2}(b_s- b_s^{\dagger})}    \,dt\,ds
\]

\[
-\int_{\mathbb{R}}\int_{\mathbb{R}}\bar{g}(t)f(s)\, e^{ \frac{K}{2}(b_s- b_s^{\dagger})}( b_s+ b_s^{\dagger})^{N-1}    e^{ \frac{k}{2}(b_t- b_t^{\dagger})}e^{ \frac{K}{2}(b_s- b_s^{\dagger})}
\]

\[
 \times ( b_t+ b_t^{\dagger})^{n-1}   e^{ \frac{k}{2}(b_t- b_t^{\dagger})}   \,dt\,ds \}
\]

Since, by lemmas \ref{ff} and \ref{fff},

\[
e^{\frac{K}{2}(b_s- b_s^{\dagger})}(b_t+ b_t^{\dagger})^{n-1}  =
\]

\[
\sum_{m=0}^{n-1}\binom{n-1}{m}(b_t+ b_t^{\dagger})^{m}K^{n-1-m}\delta^{n-1-m}(t-s)\,e^{\frac{K}{2}(b_s- b_s^{\dagger})}
\]

and

\[
  e^{\frac{k}{2}(b_t- b_t^{\dagger})}(b_s+ b_s^{\dagger})^{N-1}=
\]

\[
\sum_{m=0}^{N-1}\binom{N-1}{m}(b_s+ b_s^{\dagger})^{m}k^{N-1-m}\delta^{N-1-m}(t-s)\,e^{\frac{k}{2}(b_t- b_t^{\dagger})}
\]

and

\[
(b_t+ b_t^{\dagger})^{n-1}e^{\frac{K}{2}(b_s- b_s^{\dagger})}  =
\]

\[
e^{\frac{K}{2}(b_s- b_s^{\dagger})}\,\sum_{m=0}^{n-1}\binom{n-1}{m}(b_t+ b_t^{\dagger})^{m}K^{n-1-m}(-1)^{n-1-m}\delta^{n-1-m}(t-s)
\]

and

\[
 (b_s+ b_s^{\dagger})^{N-1}e^{\frac{k}{2}(b_t- b_t^{\dagger})}  =
\]

\[
e^{\frac{k}{2}(b_t- b_t^{\dagger})} \,\sum_{m=0}^{N-1}\binom{N-1}{m}(b_s+ b_s^{\dagger})^{m}k^{N-1-m}(-1)^{N-1-m}\delta^{N-1-m}(t-s)
\]

we find that

\[
[\hat{B}^n_k(\bar g),\hat{B}^N_K(f) ]=\frac{ 1  }{ 2^{n+N-2}}\{\sum_{m_1=0}^{n-1}\sum_{m_2=0}^{N-1}\binom{n-1}{m_1}\binom{N-1}{m_2}
\]

\[
\times(-1)^{n-1-m_1}K^{n-1-m_1}k^{N-1-m_2}
\]

\[
\times\int_{\mathbb{R}}\int_{\mathbb{R}}\bar{g}(t)f(s)\,e^{\frac{k}{2}(b_t- b_t^{\dagger})}e^{\frac{K}{2}(b_s- b_s^{\dagger})}
\]

\[
\times(b_t+ b_t^{\dagger})^{m_1} (b_s+ b_s^{\dagger})^{m_2} e^{\frac{k}{2}(b_t- b_t^{\dagger})}e^{\frac{K}{2}(b_s- b_s^{\dagger})} 
\]

\[
\times \delta^{n-1-m_1}(t-s) \delta^{N-1-m_2}(t-s)\,dt\,ds
\]

\[
-\sum_{m_3=0}^{N-1}\sum_{m_4=0}^{n-1}\binom{N-1}{m_3}\binom{n-1}{m_4}(-1)^{N-1-m_3}k^{N-1-m_3}K^{n-1-m_4}
\]

\[
\times\int_{\mathbb{R}}\int_{\mathbb{R}}\bar{g}(t)f(s)\,e^{\frac{K}{2}(b_s- b_s^{\dagger})}e^{\frac{k}{2}(b_t- b_t^{\dagger})}
\]

\[
\times(b_s+ b_s^{\dagger})^{m_3} (b_t+ b_t^{\dagger})^{m_4} e^{\frac{K}{2}(b_s- b_s^{\dagger})}e^{\frac{k}{2}(b_t- b_t^{\dagger})}
\]

\[
\times \delta^{N-1-m_3}(t-s) \delta^{n-1-m_4}(t-s)\,dt\,ds \}
\]

The case ($m_1=n-1$ , $m_2=N-1$) cancels out with ($m_3=N-1$ , $m_4=n-1$).  By the renormalization prescription (\ref{f3}) and the choice of test functions that vanish at zero, the terms $\sum_{m_1=0}^{n-3}\sum_{m_2=0}^{N-3}$ and $ \sum_{m_3=0}^{N-3}\sum_{m_4=0}^{n-3}$
are equal to zero. The only surviving terms are ($m_1=n-1$ , $m_2=N-2$), ($m_1=n-2$ , $m_2=N-1$) , ($m_3=N-1$ , $m_4=n-2$) and ($m_3=N-2$ , $m_4=n-1$) and we obtain

\[
[\hat{B}^n_k(\bar g),\hat{B}^N_K(f) ]=
\]

\[
=\frac{1}{ 2^{n+N-2}}((N-1)k-(n-1)K-(n-1)K+(N-1)k)
\]

\[
\times \int_{\mathbb{R}}\,\bar{g}(t)f(t)\,e^{ \frac{k+K}{2}(b_t- b_t^{\dagger})}( b_t+ b_t^{\dagger})^{n+N-3}  e^{ \frac{k+K}{2}(b_t- b_t^{\dagger})}\,dt
\]

\[
=\frac{2}{ 2^{n+N-2}}((N-1)k-(n-1)K)
\]

\[
\times \int_{\mathbb{R}}\,\bar{g}(t)f(t)\,e^{ \frac{k+K}{2}(b_t- b_t^{\dagger})}( b_t+ b_t^{\dagger})^{n+N-3}    e^{ \frac{k+K}{2}(b_t- b_t^{\dagger})}\,dt
\]

\[
=\frac{1}{ 2^{n+N-3}}((N-1)k-(n-1)K)
\]

\[
\times \int_{\mathbb{R}}\,\bar{g}(t)f(t)\,e^{ \frac{k+K}{2}(b_t- b_t^{\dagger})}( b_t+ b_t^{\dagger})^{n+N-3}    e^{ \frac{k+K}{2}(b_t- b_t^{\dagger})}\,dt
\]

\[
=(k(N-1)-K(n-1))\hat{B}^{n+N-2}_{k+K}(\bar g f)
\]

The proof of (\ref{viras})   follows from (\ref{wcom}) by letting $n=N=2$.
\end{proof}


\begin{thebibliography}{99}

\bibitem{1}
L. Accardi , A. Boukas, U. Franz,  Renormalized powers of quantum white noise, \emph{Infinite Dimensional Anal. Quantum Probab. Related Topics} Vol. 9, No. 1, p.129-147 (2006).

\bibitem{2}
---------, Higher Powers of $q$-deformed White Noise, to  
  appear in \emph{ Methods of Functional Analysis and Topology} (2006).

\bibitem{4}
L. Accardi , Y. G. Lu, I. V. Volovich, White noise approach to classical and quantum stochastic calculi, \emph{Lecture Notes of the Volterra International School of the same title}, Trento, Italy, 1999, Volterra Center preprint 375, Universit\`{a} di Roma Torvergata.

\bibitem{4a}
 K. Akhoumach, E. H. El Kinani, Generalized Clifford algebras and certain infinite dimensional Lie algebras,  \emph{Advances in Applied Clifford Algebras} 10 No. 1, 1-16 (2000).

\bibitem{4b}
 I. Bakas,  E.B. Kiritsis, Structure and representations of the $W_{\infty}$ algebra, \emph{Prog. Theor. Phys. Supp.} 102 (1991) 15.

\bibitem{4c}
--------, Bosonic realization of a universal W-algebra and $Z_{\infty}$ parafermions, \emph{Nucl. Phys.} B343 (1990) 185.

\bibitem{4d}
 E. H. El Kinani,  M. Zakkari, On the q-deformation of certain infinite dimensional Lie algebras, \emph{International Center for Theoretical Physics}, IC/95/163.

\bibitem{fein}
P. J. Feinsilver, R. Schott , \emph{Algebraic structures and operator calculus. Volume I: Representations and probability theory}, Kluwer 1993.

\bibitem{GD}
I. M. Gel'fand, L. A. Dikii, A family of Hamiltonian structures connected with integrable non-linear differential equations, IPM AN USSR preprint, Moscow, 1978; in \emph{Gel'fand Collected Papers}, edited by Gindinkin et al., Springer-Verlag N.Y, 1987, p. 625


\bibitem{5}
 K. Ito , \emph{On stochastic differential equations},
Memoirs Amer. Math. Soc. 4 (1951).

\bibitem{ketov}
S. V. Ketov , \emph{Conformal field theory}, World Scientific, 1995.

\bibitem{6}
K. R. Parthasarathy , \emph{ An introduction to quantum stochastic calculus}, Birkhauser Boston Inc., 1992.

\bibitem{7}
C. N. Pope , Lectures on W algebras and W gravity, \emph{Lectures given at the Trieste Summer School in High-Energy Physics}, August 1991.


\bibitem{8}
A.B. Zamolodchikov ,  Infinite additional symmetries in two-dimensional conformal quantum field theory, \emph{Teo. Mat. Fiz.} 65 (1985), 347-359. 

\end{thebibliography}
\end{document}